\documentclass[aps,prl,twocolumn,longbibliography,superscriptaddress, nobibnotes, 10pt,floatfix]{revtex4-2}
\usepackage[utf8]{inputenc}
\usepackage[T1]{fontenc}
\usepackage{amsmath,amssymb}
\usepackage{xcolor}
\usepackage{physics}
\usepackage[normalem]{ulem}
\usepackage{xspace}
\usepackage{comment}
\usepackage[colorlinks=true,
            linkcolor=blue,
            urlcolor=cyan,
            citecolor=cyan,
            filecolor=magenta]{hyperref}
\usepackage{mathtools}
\usepackage{graphicx}
\usepackage{dcolumn}
\usepackage{bm}
\usepackage{dsfont}
\usepackage{orcidlink}
\definecolor{Salmon}{RGB}{250,128,114}

\usepackage[normalem]{ulem}

\newcommand{\Str}{\mathrm{Str}}

\begin{document}


\newcommand{\getZuerichAffiliation}{\affiliation{Institut f{\"u}r Theoretische Physik, ETH Z{\"u}rich, Wolfgang-Pauli-Str. 27, 8093 Z{\"u}rich, Switzerland}}

\newcommand{\getHarvardChemAffiliation}{\affiliation{Department of Chemistry and Chemical Biology, Harvard University, Cambridge, Massachusetts 02138, USA}}


\title{Emergent quantum chaos from correlations on a random graph}
\author{Mrinal Sarkar\,\orcidlink{0000-0003-3112-6530}}
\thanks{Corresponding author}\email{sarkar@thphys.uni-heidelberg.de}
\affiliation{Institut für Theoretische Physik, Universität Heidelberg, 69120 Heidelberg, Germany 
}
\author{Valerio Pagni\,\orcidlink{0009-0007-5323-4651}}\getZuerichAffiliation\getHarvardChemAffiliation
\author{Tilman Enss\,\orcidlink{0000-0002-5334-2448}}%
\affiliation{Institut für Theoretische Physik, Universität Heidelberg, 69120 Heidelberg, Germany 
}%
\author{Nicolò Defenu\,\orcidlink{0000-0002-3401-3665}}%
\getZuerichAffiliation

\begin{abstract}
This work demonstrates that sparse long-range random bonds on a one dimensional lattice alone can generate quantum-chaotic spectral correlations and also drive a localization transition in a noninteracting single-particle Hamiltonian. The model is a one-dimensional ring in which each pair of sites is connected independently with a probability $p_{ij}= d_{ij}^{-(1+\sigma)}$. Each bond carries identical unit hopping and on-site disorder is absent. Despite the absence of on-site disorder and interaction, the model displays quantum chaotic spectra with Gaussian orthogonal ensemble (GOE) level statistics at small $\sigma$ and localized eigenstates with Poisson statistics at larger $\sigma$. The transition occurs in the range $ 0.80 \lesssim \sigma_c \lesssim 0.85$, far above the summability threshold of the mean hopping profile ($\sigma=0$). A Gaussian field theory retaining only the mean and variance of the Bernoulli bonds instead predicts a threshold at $\sigma=1$, suggesting that higher cumulants are infrared-relevant. Our findings hint towards a universality class that is distinct from both the power-law random banded matrix model and the standard Anderson transition.
\end{abstract}

\maketitle

\emph{Introduction}---The emergence of  ergodic behavior and its breakdown by microscopic disorder remains one of the central themes of condensed matter physics, linking single-particle transport to many-body chaos~\cite{landau2013statistical, d2016quantum, haake1991quantum, srednicki1994chaos, deutsch1991quantum, nandkishore2015many}. Since Anderson's seminal work~\cite{anderson1958absence}, disorder-induced localization has been extensively studied in noninteracting quantum systems~\cite{ohtsuki1999review, evers2008anderson, kravtsov2015random, tikhonov2016anderson, tikhonov2016fractality, tarquini2017critical, garcia2022critical, vanoni2024renormalization, altshuler2025renormalization}. In the standard Anderson model, a particle hops on a lattice with random on-site potentials, and destructive interference between multiply scattered paths suppresses diffusion~\cite{evers2008anderson}. Scaling theory predicts that, for short-range hopping, all single-particle states are localized in $d\leq2$ for arbitrarily weak disorder~\cite{abrahams1979scaling}. This paradigm can be circumvented by Hamiltonians with long-range (LR) structure, where hopping amplitudes or disorder correlations decay algebraically with distance. Such LR systems admit extended single-particle states in one dimension, restoring ergodicity and enabling localization-delocalization transitions that are absent in short-range (SR) models~\cite{levitov1989absence,levitov1990delocalization, rodriguez2000quantum,rodriguez2003anderson,malyshev2004monitoring,de2005localization, mirlin1996transition, deMoura1998deloca}.

One route is to consider an Anderson model with deterministic hopping amplitudes $t_{ij} \sim d_{ij}^{-\alpha}$ ($d_{ij}$ being the distance between the sites $i$ and $j$) and uncorrelated diagonal disorder. For $\alpha<1$, the range of the kinetic term is long enough to support extended states~\cite{levitov1989absence,levitov1990delocalization}, while for $\alpha>1$, extended states are restricted to narrow windows near a band edge ~\cite{rodriguez2000quantum,rodriguez2003anderson,malyshev2004monitoring,de2005localization}.
Another paradigm is the power-law random banded matrix (PRBM) ensemble, in which the hopping amplitudes between sites $i$ and $j$ ($i \neq j$) are uncorrelated and Gaussian distributed with zero mean and variance
$\langle |H_{ij}|^2\rangle \sim |i-j|^{-2\alpha}$~\cite{mirlin1996transition}. The PRBM ensemble exhibits extended states for $\alpha<1$, localized states for $\alpha>1$, and a critical point at $\alpha=1$ with multifractal eigenstates~\cite{mirlin1996transition, kravtsov1996spectral}. LR correlations in the on-site disorder can also drive localization-delocalization transitions in an otherwise fully localized 1D chain\,\cite{deMoura1998deloca}. 

\begin{figure}[tbp!]
\centering
\includegraphics[width=0.99\linewidth]{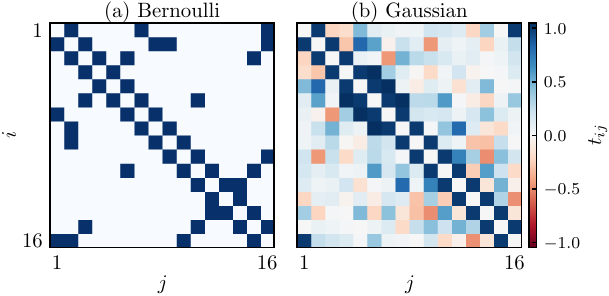}
\caption{Hopping matrices for a typical disorder realization of the 1DLR3 graph with $L=16$ sites and $ \sigma=0.6$. \textbf{(a)} Bernoulli disorder: Each off-diagonal element $t_{ij}$ is $1$ (Blue) with probability $p_{ij} = d_{ij}^{-(1+\sigma)}$ and $0$ (White) otherwise, yielding a sparse binary matrix. \textbf{(b)} Gaussian truncation: The matrix elements $t_{ij}$ are replaced by a Gaussian with the same mean $p_{ij}$ and variance $p_{ij} (1 - p_{ij})$, truncating at the second cumulant of the Bernoulli distribution. The resulting hopping matrix is dense and continuously distributed.}
\label{fig:hopping_matrix}
\end{figure}

All the aforementioned cases share an important feature: the underlying geometry is fixed, and  disorder is additive to the geometry. A natural question then arises: What happens when the disorder is purely geometrical and affects solely the kinetic pathways available for propagation? Does interference still affect the thermodynamic behavior of the system also in these complex geometries?

We address these questions in a minimal setting where disorder and connectivity are inseparable. We consider a one-dimensional long-range random ring (1DLR3): a ring of $L$ sites in which each pair of sites $(i,j)$ is connected independently with probability
$p_{ij}= d_{ij}^{-\alpha}$, with $\alpha=1+\sigma$. Every occupied bond carries unit hopping amplitude and onsite disorder is absent. The resulting Hamiltonian is therefore neither an Anderson model with an added random potential, nor a random-matrix ensemble with independently distributed hopping amplitudes~\cite{levitov1989absence,levitov1990delocalization, rodriguez2000quantum,rodriguez2003anderson,malyshev2004monitoring,de2005localization, mirlin1996transition, deMoura1998deloca, deng2018Duality, nosov2019corr, kutlin2020renormalization}. Its quenched randomness is solely geometric: the presence of sparse random LR bonds, see Fig.~\ref{fig:hopping_matrix}. These bonds play a dual role, simultaneously defining the kinetic network on which the particle propagates and generating the scattering responsible for localization.

A heuristic unifying criterion for ergodic-to-localization threshold in the aforementioned LR models is provided by the summability of the effective hopping profile. In 1D, a power-law profile $t_{ij}\sim d_{ij}^{-\alpha}$ is non-summable for $\alpha \leq 1$ and summable for $\alpha>1$. This heuristic is placed on microscopic footing by Levitov's resonance criterion for strongly disordered power-law hopping models, identifying $\alpha=d$ as the threshold in $d$ dimensions~\cite{levitov1989absence,levitov1990delocalization}. Although the PRBM ensemble lies outside Levitov's framework, its typical off-diagonal matrix elements also scale as $|H_{ij}|_{\rm typ}\sim |i-j|^{-\alpha}$, and the transition occurs at the same value~\cite{mirlin1996transition}. Thus, the threshold $\alpha=1$ separates a genuinely (non-summable) LR regime, where extended states can be sustained, from a summable regime, where localization is generally expected.

We show that the 1DLR3 graph exhibits two related surprises. First, purely geometric and sparse disorder is indeed sufficient to generate quantum-chaotic spectral correlations. For sufficiently slowly decaying bond probabilities, the level statistics approach those of the Gaussian orthogonal ensemble (GOE), the hallmark of ergodic (equivalently, extended or chaotic) behavior in single-particle systems. Second, and more unexpectedly, this chaotic phase emerges already for relatively weak long-range correlations at a threshold lying in the range $1.80 \lesssim \alpha_c \lesssim 1.85$ (equivalently, $0.80 \lesssim \sigma_c \lesssim 0.85$). The location of the threshold is striking because it is far beyond the naive summability threshold $\alpha=1$ {($\sigma=0$)} set by replacing the random adjacency matrix with its mean hopping $\overline{t}_{ij}=p_{ij}= d_{ij}^{-(1+\sigma)}$. We characterize the resulting transition through large-scale exact diagonalization, computing spectral and eigenstate observables, supplemented by an effective field-theory perspective. The latter suggests that a Gaussian description of the bond disorder is incomplete: the sparse Bernoulli structure of the graph, and in particular its non-Gaussian higher cumulants, play a central role in determining the precise location of the transition.

\begin{figure}[tbp!]
\centering
\includegraphics[width=0.95\linewidth]{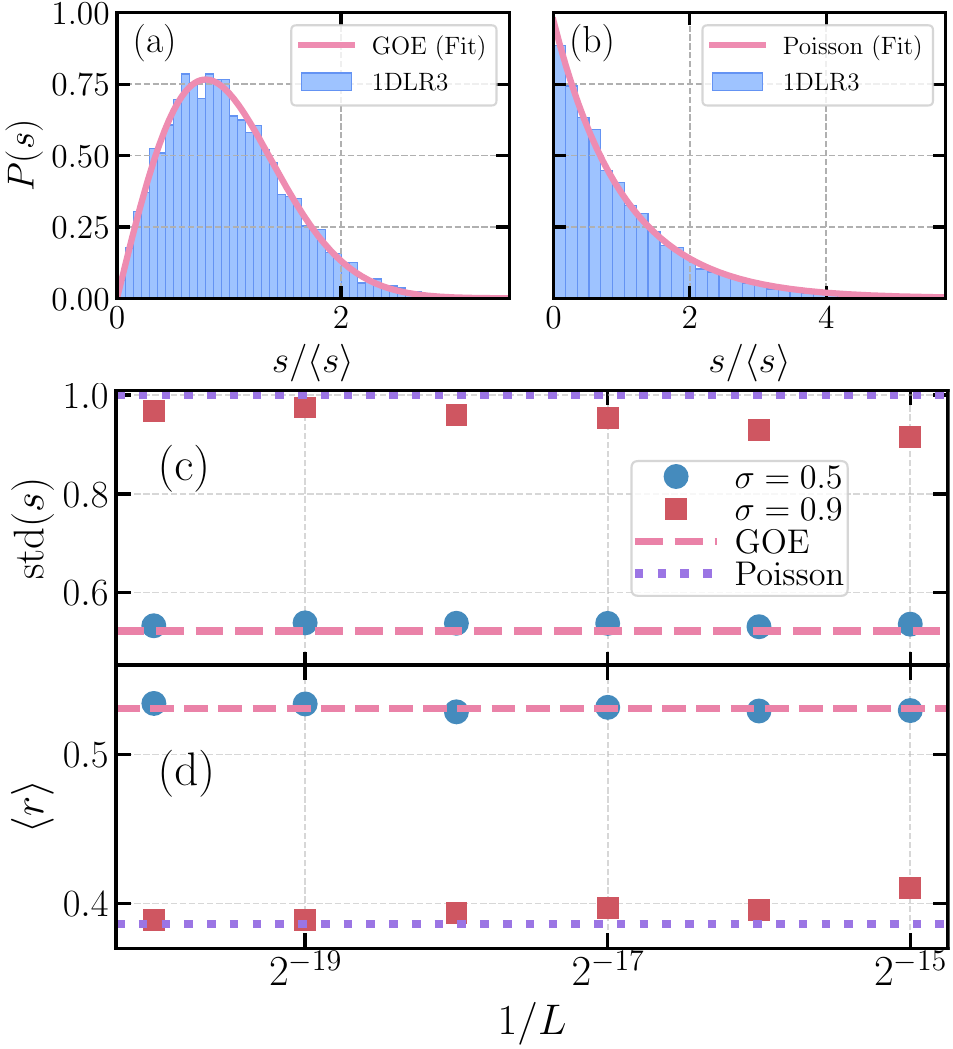}
\caption{Eigenvalue statistics for $\sigma = 0.5$ [panels(a), (c), (d)] and $\sigma = 0.9$ [panels (b), (c) and (d)]. Panels (a) and (b) show the level-spacing distribution $P(s)$ for $L = 2^{20}$. In panel~(a) ($\sigma = 0.5$), $P(s)$ follows the Wigner surmise for the Gaussian Orthogonal Ensemble (GOE); a fit to $f(s) = 2\,B\,s\,e^{-Bs^{2}}$ yields $B = 0.804 \pm 0.012$ ($\approx \pi/4$), in excellent agreement with the theoretical GOE prediction. The fitted curve is shown by the magenta line. In panel~(b) ($\sigma = 0.9$), $P(s)$ is well described by the Poisson form $f(s) = \lambda e^{-\lambda s}$ with $\lambda = 0.97(1)$ (magenta line); the absence of level repulsion is consistent with localized eigenstates. Panel (c) shows $\mathrm{std}(s)$ as a function of inverse system size $1/L$ for $\sigma = 0.5$ (Blue circles) and $\sigma = 0.9$ (Red squares). For all $L$ values under study, $\mathrm{std}(s)$ is nearly saturated to the GOE value $\sqrt{4/\pi - 1} \approx 0.523$ for $\sigma = 0.5$, whereas it approaches the Poisson value $\mathrm{std}(s) = 1$ with increasing $L$ for $\sigma = 0.9$. Panel (d) shows the scaling of the mean gap ratio $\langle r \rangle$ with system-size $1/L$, which saturates to $\langle r \rangle_{\mathrm{GOE}} \approx 0.5307$ and $\langle r \rangle_{\mathrm{P}} \approx 0.3863$, respectively. These results consistently indicate an ergodic phase at $\sigma = 0.5$ and a localized phase at $\sigma = 0.9$.}
\label{fig:evalue_stat}
\end{figure}

\emph{The model}---We consider the transport of a single quantum particle on a 1DLR3 graph. The graph is constructed as follows: Consider first a 1D lattice of $L$ sites arranged on a ring. A bond between two sites $i,j \in \{ 1, \ldots, L\}$ is added with probability $p_{ij} = d_{ij}^{-(1+\sigma)}$, where the distance is measured along the backbone $d_{ij} = {\sin\bigl(\pi|i-j|/L\bigr)}/{\sin(\pi/L)}$~\cite{millan2021complex,sarkar2024universality}. Note that the distance reduces to $|i-j|$ for short separations and saturates at order $L$ for antipodal sites. The parameter $\sigma>0$ controls the connectivity of the graph. A single quantum particle on this graph is described by the tight-binding Hamiltonian 
\begin{align}
    H= - \sum_{i < j} t_{ij}\,(c_i^{\dagger} c_j+\textrm{h.c.}), 
    \label{eq:Ham}
\end{align}
where the hopping amplitude $t_{ij} = 1 $ with probability $p_{ij}$ and $t_{ij} = 0 $ with probability $1 -p_{ij}$. The nearest-neighbor bonds are therefore always present, while longer bonds are added with algebraically decaying probability, see the depiction of the $t_{ij}$ matrix in the left panel of Fig.~\ref{fig:hopping_matrix}.

\begin{figure}[tbp!]
\centering
\includegraphics[width=0.99\linewidth]{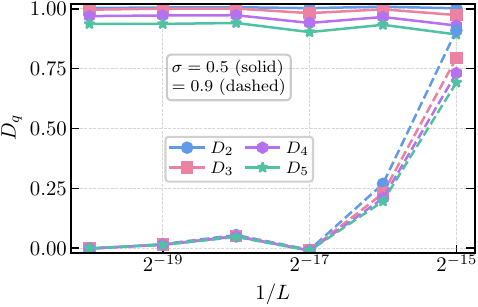}
\caption{Scaling of the fractal dimensions $D_q$ for $q = 2, 3, 4, 5$ with inverse system size $1/L$. Solid lines show results for $\sigma = 0.5$, where $D_q$ tends to approach unity
as $L \to \infty$, consistent with the GOE-like level statistics observed in Fig.\,\ref{fig:evalue_stat}, yielding an existence of ergodic phase. Dashed lines show results for $\sigma = 0.9$, where $D_q$ extrapolates to zero in the thermodynamic limit, confirming localization consistently with the level statistics.}
\label{fig:estate_stat}
\end{figure}

A central feature of this model is its Bernoulli, and hence intrinsically non-Gaussian, disorder. The mean hopping is $\overline t_{ij}=p_{ij}= d_{ij}^{-(1+\sigma)}$, and the variance has the same asymptotic decay,
\begin{equation}
    C_{ij}
    \equiv
    \overline{\delta t_{ij}^{\,2}}
    =
    p_{ij}(1-p_{ij})
    \sim d_{ij}^{-(1+\sigma)}
    \qquad (d_{ij}\gg 1).
    \label{eq:bernoulli_variance}
\end{equation}
More generally, since $t_{ij}\in\{0,1\}$, all higher Bernoulli cumulants share the same leading long-distance tail. Thus the first two moments naturally suggest a connection to dense power-law random-matrix ensembles, while the graph retains non-Gaussian structure at long distances.

\emph{Numerical results}---For system sizes of up to $L= 2^{20}$, the computation of the entire spectrum by exact diagonalization is prohibitive. We instead target $n_{\mathrm ev} = 100$ eigenstates near the band center $E^*=0$ using the shift-invert Lanczos method~\cite{schenk2008large}, averaging over $100$ independent realizations of the random graph.

The transport properties are probed through two different, complementary approaches. First, spectral statistics~\cite{bohigas84}: To characterize level correlations, we define the adjacent level spacing $s_n = E_{n+1} - E_n$ for an ordered spectrum $\{ E_n \}$ and compute the consecutive-gap ratio
\begin{equation}
  r_n = \frac{\min(s_n, s_{n-1})}{\max(s_n, s_{n-1})}
  \label{eq:r_ratio}
\end{equation}
and its mean $\langle r \rangle$ on all available levels and graph realizations.  The Wigner-Dyson (GOE) value $\langle r \rangle_{\rm GOE} \approx 0.5307$ signals level repulsion characteristic of an ergodic metallic phase, while the Poisson value $\langle r \rangle_{\rm Poisson} \approx 0.3863$ signals level clustering consistent with localization. From the unfolded spectrum we also compute the nearest-neighbor level-spacing distribution $P(s)$, where $s$ denotes the normalized spacing between consecutive eigenvalues. The resulting distribution is compared with the limiting predictions of random-matrix theory. In the localized phase, level spacings obey Poisson statistics, $P(s) = e^{-s}$, whereas in the non-localized phase, they follow the Gaussian Orthogonal Ensemble (GOE) Wigner surmise, $P(s) = \frac{\pi}{2}s\,\exp\left(-\pi s^2/4\right)$. The crossover between these two distributions serves as a spectral diagnostic of the localization transition.

\begin{figure}[tbp!]
\centering
\includegraphics[width=0.99\linewidth]{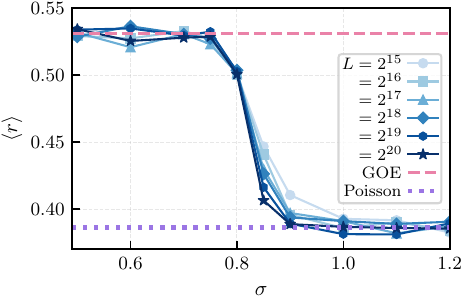}
\caption{Mean gap ratio $\langle r \rangle$ as a function of $\sigma$ for $L = 2^{15}$ to $2^{20}$. The curves cross over from the GOE value $\langle r \rangle_{\mathrm{GOE}} \approx 0.5307$ at small $\sigma$ to the Poisson value $\langle r \rangle_{\mathrm{P}} \approx 0.3863$ at large $\sigma$, indicating an ergodic-to-localization transition between $\sigma = 0.80$ and $0.85$.}
\label{fig:r_vs_sigma}
\end{figure}

Figure~\ref{fig:evalue_stat} shows the eigenvalue statistics for two representative values of $\sigma$. At $\sigma = 0.5$, the level-spacing distribution $P(s)$ follows the Wigner surmise for the GOE and the mean gap ratio $\langle r \rangle$ is nearly saturated to $\langle r \rangle_{\mathrm{GOE}} \approx 0.5307$ for all $L$ studied, signaling an ergodic phase. At $\sigma = 0.9$, $P(s)$ follows Poisson statistics and $\langle r \rangle$ approaches $ \langle r \rangle_{\mathrm{P}} \approx 0.3863$ with an increase in $L$, consistent with localization.

A second diagnostic tool is provided by eigenstate statistics~\cite{evers2008anderson}. For each normalized eigenstate $|n\rangle$, we define the generalized inverse participation ratio (IPR) $I_q = \sum_{i} |\langle i | n \rangle|^{2q}$. Normalization corresponds to $q=1$, while $q=2$ corresponds to the conventional IPR. Studying a range of $q$ values allows us to distinguish ergodic, localized and multifractal states. Here we focus on $q > 1$; larger values of $q$ amplify wavefunction peaks and probe the fraction of sites with dominant support, providing complementary information beyond $q = 2$ alone.
In general, for large $L$, the moments scale as $I(q) \sim L^{- \tau(q)}$, where $D_q= \tau(q)/(q-1)$ is called the fractal dimension. In an extended ergodic state, $D_q = 1$, while for a localized state $D_q=0 \;\forall q>1$. Multifractal, i.e., extended but non-ergodic states exhibit a non-trivial $q$-dependence of $D_q$, with intermediate scaling $0 < D_q <1$.

To further confirm the localized phase at $\sigma = 0.9$, we compute the fractal dimensions $D_q$ for $q = 2, 3, 4, 5$. The scaling of $D_q$ with $1/L$ shown in Fig.\,\ref{fig:estate_stat}(dashed lines) extrapolates to $D_q \to 0$ in the thermodynamic limit, consistent with the Poisson level statistics observed above. At $\sigma = 0.5$,  $D_q$ tends to approach unity asymptotically for all $q$-values [Fig.\,\ref{fig:estate_stat}(solid lines)], indicating an ergodic phase. This implies that the system undergoes an ergodic-to-localized transition as $\sigma$ increases. Although pinpointing the exact transition point is not our primary goal here, our data indicate that the threshold lies in the range $0.80 \lesssim \sigma_c \lesssim 0.85$ (see Fig.~\ref{fig:r_vs_sigma}), clearly excluding a transition at $\sigma=1$.

\emph{Effective theory: Gaussian truncation and its limits}---We now ask what analytical expectations follow from a simple disorder-averaged field-theoretical description of the 1DLR3. The central quantity is the single-particle Green's function $G^{R/A}(E) = {\left( E - H \pm i0\right )}^{-1}$, where $R$ and $A$ stand for retarded and advanced, respectively, and $H$ is the Hamiltonian for a given disorder realization. Physical observables such as density of states and level statistics are expressed in terms of disorder-averaged products of $G^R$ and $G^A$. Since the hopping variables are Bernoulli distributed, an exact disorder average generates cumulants of all orders. As a first approximation, we retain only the mean hopping profile $\overline t_{ij}=p_{ij}$ and the second cumulant $C_{ij}$, see Eq.~\eqref{eq:bernoulli_variance}. This Gaussian truncation discards all effects of the discrete $\{ 0,1\}$ nature of the hopping, see Fig.\,\ref{fig:hopping_matrix}.

Within the Gaussian truncation, one can proceed to the supersymmetric construction of the nonlinear sigma
model (NL$\sigma$M)~\cite{efetov1983supersymmetry,efetov1999supersymmetry,mirlin1996transition,evers2008anderson}. We introduce, at each site, a supervector $\Psi_i$ whose bosonic (commuting) components generate the Green's functions $G^{R/A}$ as two-point correlators, while its fermionic  (anticommuting) components ensure unit normalization of the field integral for each disorder realization. This allows disorder averages of products of $G^R$ and $G^A$ to be computed as Gaussian integrals over $\Psi_i$. The generating functional before disorder averaging is
\begin{equation}
    Z = \int D(\bar\Psi,\Psi)\, e^{i\sum_{ij}
    \bar\Psi_i \mathcal{H}_{ij} \Psi_j},
    \label{eq:susy_generating_function}
\end{equation}
where the supermatrix $\mathcal{H}_{ij}$ is
\begin{equation}
    \mathcal{H}_{ij} = \left(E + \frac{\omega+i0}{2}\Lambda \right)\delta_{ij} - H_{ij},
    \,\,
    \Lambda= \begin{pmatrix}
        \mathds{1} & 0 \\
        0 & -\mathds{1}
    \end{pmatrix}_{RA}.
\end{equation}
The matrix $\Lambda$ acts in retarded/advanced space with eigenvalues $+1$ ($R$) and $-1$ ($A$). The frequency $\omega$ is a small symmetry-breaking parameter that distinguishes $G^R$ from $G^A$.

To perform disorder averaging, split the Hamiltonian as $H = \bar{H} + \delta H$, where $\bar{H}_{ij} = \bar{t}_{ij}$ is the deterministic mean hopping and $\delta H_{ij} = \delta t_{ij}$ is the random disorder. Under Gaussian truncation, the latter yields on disorder averaging a quartic interaction between local supermatrix bilinears $B_i=\Psi_i\bar\Psi_i$,
\begin{equation}
S_{\rm int} = \frac{1}{2} \sum_{ij} C_{ij}\, \Str(B_iB_j),
\label{eq:quartic_interaction}
\end{equation}
where $\Str$ denotes the supertrace. We then decouple Eq.~\eqref{eq:quartic_interaction} by introducing a Hubbard--Stratonovich supermatrix field $R_i$ conjugate to $B_i$. Integrating out the supervectors gives
\begin{equation}
\begin{aligned}
S_{\rm HS}[R] = \frac{1}{2} \sum_{ij} (C^{-1})_{ij}
\Str(R_iR_j) + \Str' \ln M_{ij}[R],
\end{aligned}
\label{eq:HS_action}
\end{equation}
where $\Str'$ is the supertrace together with the trace over lattice sites and  $M_{ij}[R] = \left( E+\frac{\omega+i0}{2}\Lambda-R_i \right)\delta_{ij} - \overline t_{ij}$.

The disorder-averaged functional $\langle Z\rangle = \int DR\,e^{-S_{\rm HS}[R]}$ is then evaluated in the saddle-point approximation. Setting $\delta S_{\rm HS}/\delta R_i = 0$ yields the self-consistent Born approximation~\cite{altland2010condensed}
\begin{equation}
R_i = \sum_j C_{ij}\,G_{jj}[R],
\label{eq:saddle_point}
\end{equation}
where $G[R] = M[R]^{-1}$ is the single-particle Green's function dressed by the self-energy $R_i$. Equation\,\eqref{eq:saddle_point} admits a translationally invariant solution $R_i = R_{\star} = \Sigma'\mathds{1} -i\gamma\Lambda$, where $\Sigma'$ is disorder-induced energy shift and $\gamma >0$ is the elastic scattering rate. The latter is related to the density of states by $\nu (E) = \gamma/C(0)$, see End Matter.

The uniform saddle $R_\star$ is not unique. At $\omega=0$, $S_{\rm HS}$ in Eq.~\eqref{eq:HS_action} is invariant under global superspace rotations $R_\star \to T^{-1}R_\star T$, generating a manifold of degenerate saddles $R_Q = \Sigma'\mathbf{1} - i\gamma Q$, where $Q = T^{-1}\Lambda T$ and $Q^2=\mathbf{1}$. Promoting $T$ to a slowly varying field $T_i$ yields soft fluctuation modes $Q_i = T_i^{-1}\Lambda T_i$. Physically these diffusion modes control the long-distance transport.

To quadratic order in the soft modes one obtains a nonlocal NL$\sigma$M
\begin{equation}
\begin{aligned}
S[Q] &= - \frac{\gamma^2}{2N} \sum_q \mathcal K(q)\, \Str(Q_qQ_{-q})\\
&\quad - i \frac{\pi\nu(E)}{2} (\omega+i0) \sum_i \Str(\Lambda Q_i).
\end{aligned}
\label{eq:LR3_NLSM_general}
\end{equation}
Here, the stiffness $\mathcal K(q)$, controlling the cost of spatial fluctuations, can be written in manifestly massless form
\begin{equation}
\mathcal K(q) = \left[ C^{-1}(q)-C^{-1}(0) \right] +
\left[ \Pi(0)-\Pi(q) \right],
\label{eq:LR3_kernel_split}
\end{equation}
where $\Pi(q) = \int_k G_\star^R(k+q/2)\,G_\star^A(k-q/2)$ is the particle-hole bubble, measuring the propagation of a particle-hole pair of momentum $q$ through the effective medium. The first bracket in Eq.~\eqref{eq:LR3_kernel_split} is generated by the variance of the random hopping, while the second is the correction due to the non-zero mean hopping $\overline{t}_{ij}$.

Since, in our case, $C_{ij} \sim \overline{t}_{ij} = d_{ij}^{-(1+\sigma)}$ the variance contribution gives
\begin{equation}
C^{-1}(q)-C^{-1}(0)=A_\sigma |q|^\sigma+\cdots , \qquad A_\sigma>0 .
\label{eq:variance_kernel}
\end{equation}
while the mean hopping produces a separate infrared correction, yielding
\begin{equation}
\Pi(0)-\Pi(q) \sim
\begin{cases}
|q|^{1+2\sigma}, & 0<\sigma<1/2,\\
q^2, & \sigma>1/2,
\end{cases}
\label{eq:deterministic_kernel}
\end{equation}
up to logarithmic corrections at $\sigma=1/2$. Thus the contribution of mean hopping is subleading compared with the variance-generated $|q|^\sigma$ term ($0 < \sigma < 2$).

The critical point between extended and localized phases in this model is found in correspondence of the scaling $\mathcal{K}(q) \sim |q|$, as described in Ref.~\cite{mirlin1996transition,evers2008anderson}. If one retained a sufficiently short-ranged hopping variance, the leading infrared stiffness would be set by the bubble contribution Eq.~\eqref{eq:deterministic_kernel}, placing the critical threshold at $\sigma_c^{(1)}=0$. This would be consistent with the naive summability criterion. However, the Bernoulli variance changes this conclusion qualitatively. As observed in Eq.~\eqref{eq:variance_kernel}, under Gaussian truncation, the variance produces the more singular kernel $|q|^\sigma$, dominating over the bubble at small $q$. This shifts the threshold to $\sigma_c^{(2)}=1$.

However, our numerical simulations demonstrate a transition at $0.80 \lesssim \sigma_c \lesssim 0.85$ where the Gaussian truncation predicts chaotic behavior. It points to the relevance of non-Gaussian nature of the Bernoulli disorder. Indeed, all higher cumulants have the same leading long-distance decay $d_{ij}^{-(1 +\sigma)}$, and cannot be treated as short-distance corrections to the Gaussian theory. This suggests a natural interpretation in terms of a hierarchy of approximations. The mean hopping alone places the threshold at $\sigma_c^{(1)}=0$; including the variance shifts it to $\sigma_c^{(2)}=1$; retaining higher cumulants are expected to further renormalize the Gaussian prediction by a finite amount yielding a value consistent with the numerics. Nevertheless, our study suggests that the Gaussian truncation captures the dominant effect of graph disorder beyond the mean hopping, while the remaining shift reflects the non-Gaussian sparsity of the graphs.

\emph{Conclusions}---To conclude, we have shown that sparse long-range geometry alone can generate both quantum-chaotic spectral correlations and localization in a noninteracting single-particle Hamiltonian. The location of the threshold indicates that the sparse Bernoulli structure of the graph is not a harmless microscopic detail. Since the same rare LR bonds provide the only channels for long-distance propagation, their ability to generate chaotic spectra or localization through their quenched spatial arrangement is captured by neither a simple picture of deterministic long-range hopping nor the Gaussian description of the bond fluctuations, indicating a different universality class. The mechanism is also different from recently studied structurally disordered three-dimensional lattices, where the Anderson transition is driven by a smooth tuning of geometric disorder strength and was found to belong to the standard Anderson universality class~\cite{bhattacharjee2025anderson}.

Our work opens up several promising directions including a finite-size scaling analysis of the critical properties, a field-theory treatment of the non-Gaussian vertices, and a detailed characterization of the possible nonergodic extended regime near the transition. Moreover, quantum graphs with GOE spectral statistics have recently been realized on silicon photonic chips~\cite{girin2026graphschipsiliconphotonics}, making these systems a natural experimental testbed for our findings.

\begin{acknowledgments}
This research was funded by the Swiss National Science Foundation (SNSF) grant numbers 200021--207537 and 200021--236722, by the Deutsche Forschungsgemeinschaft (DFG, German Research Foundation) under Germany's Excellence Strategy EXC2181/1-390900948 (the Heidelberg STRUCTURES Excellence Cluster) and the Swiss State Secretariat for Education, Research and Innovation (SERI). M.S. also acknowledges support by the state of Baden-Württemberg through bwHPC cluster.

\end{acknowledgments}


\bibliography{References_mrinal}

\clearpage
\appendix
\section*{End Matter}

\subsection{Derivation of the nonlinear sigma model}
\label{app:gaussian_nlsm}

In this appendix we derive Eq.  \eqref{eq:LR3_NLSM_general}, i.e. the nonlinear sigma model that retains the first two cumulants of the 1DLR3 model. We start from the Hubbard-Stratonovich action \eqref{eq:HS_action}, rewritten here as
\begin{equation}
    S_{\rm HS}
    =
    \frac{1}{2}
    \sum_{ij}
    (C^{-1})_{ij}
    \Str(R_iR_j) +
    \Str'
    \ln
    M_{ij}[R],
\label{eq:HS_rewritten}
\end{equation}
where
\begin{equation}
    M_{ij}[R]
    =
        \big(
            E+\tfrac{\omega+i0}{2}\Lambda-R_i
        \big)\delta_{ij}
        -
        \bar t_{ij}.
\end{equation}
We look for a translationally invariant saddle point, parametrized by the retarded/advanced ($R/A$) form
\begin{equation}\label{eq:saddle_point_R*}
    R_i=R_\star=\Sigma'\mathds{1}-i\gamma\Lambda ,
\end{equation}
where $\gamma>0$ inside the mean-field band. Varying Eq.~\eqref{eq:HS_rewritten} with respect to $R_i$ yields the saddle-point condition
\begin{equation}
    R_i
    =
    \sum_j C_{ij}\,G_{jj}[R],
\label{eq:app_saddle_point}
\end{equation}
where $G[R]$ is the inverse of $M[R]$. The latter corresponds to the self-consistent Born approximation for the self-energy~\cite{altland2010condensed}, thus allowing for the identification of the saddle point $R_\star$ with the self-energy due to disorder scattering, and justifying the notation in Eq.~\eqref{eq:saddle_point_R*}.
For a translationally invariant saddle, $G_{ii}$ is
independent of $i$, as well as $C(0)\equiv \sum_j C_{ij}$, and therefore
\begin{equation}
    R_\star^{R/A}
    =
    C(0)
    \int_k
    \left[
        E-\varepsilon(k)-R_\star^{R/A}
    \right]^{-1},
    \label{eq:app_saddle}
\end{equation}
with $\varepsilon(k)=\sum_r e^{-ikr}\bar t(r)$ and  $\int_k\equiv \frac{1}{N}\sum_k$. To identify the saddle manifold we have set
$\omega=0$, keeping the infinitesimal regulator implicit. A finite
frequency explicitly breaks the retarded/advanced symmetry and will be
restored later as a symmetry-breaking term.

The saddle $R_\star$ is not unique. At $\omega=0$, the action is invariant under
global rotations in superspace:
\begin{equation}
    R_Q
    =
    T^{-1}R_\star T
    =
    \Sigma'\mathds{1}-i\gamma Q,
\end{equation}
where $Q=T^{-1}\Lambda T$ and $Q^2=\mathds{1}$, is an equivalent uniform saddle. Distinct saddles are parametrized by the coset obtained by quotienting
out rotations that commute with $\Lambda$, since those leave $Q$ unchanged. The soft modes are obtained by promoting the global rotation to a slowly varying one, $T\to T_i$.

We now expand the action in the angular fluctuations around
the uniform saddle. As the first variation vanishes by construction, we consider the second variation of Eq.~\eqref{eq:HS_action}:
\begin{equation}
    S^{(2)}
    =
    \frac{1}{2}
    \sum_{ij}
    (C^{-1})_{ij}
    \Str(\delta R_i\delta R_j)
    -
    \frac{1}{2}
    \Str'
    \left(
        G_\star\delta M\,G_\star\delta M
    \right),
    \label{eq:app_second_variation}
\end{equation}
where $\delta M_{ij} = -\delta R_i\,\delta_{ij}$ and
\begin{equation}
    M_{\star,ij}
    =
    (E-R_\star)\delta_{ij}-\bar t_{ij},
    \qquad
    G_\star=M_\star^{-1}.
\end{equation}
We parametrize the angular modes as
\begin{equation}
    Q_i=e^{-W_i/2}\Lambda e^{W_i/2},
    \qquad
    \{W_i,\Lambda\}=0.
\end{equation}
To linear order, $Q_i\simeq\Lambda+\Lambda W_i$, so that $\delta R_i \simeq -i\gamma\Lambda W_i$.
The Hubbard--Stratonovich part of \eqref{eq:app_second_variation} gives
\begin{equation}
    S_C^{(2)}
    =
    \frac{\gamma^2}{2}
    \int_q
    C(q)^{-1}
    \Str(W_qW_{-q}).
\end{equation}
The logarithmic part gives
\begin{equation}
    S_{\log}^{(2)}
    =
    \frac{\gamma^2}{2}
    \int_{q,k}
    \Str
    \left[
        G_\star\!\left(k+\tfrac q2\right)
        \Lambda W_q
        G_\star\!\left(k-\tfrac q2\right)
        \Lambda W_{-q}
    \right].
    \label{eq:app_log_trace}
\end{equation}
Since $G_\star$ is diagonal in retarded/advanced space, while $W_q$ is
off-diagonal, carrying out the trace yields only retarded--advanced products:
\begin{equation}
    S_{\log}^{(2)}
    =
    -\frac{\gamma^2}{2}
    \int_q
    \Pi(q)
    \Str(W_qW_{-q}),
    \label{eq:app_log_bubble}
\end{equation}
with
\begin{equation}
    \Pi(q)
    =
    \int_k
    G_\star^R\!\left(k+\frac q2\right)
    G_\star^A\!\left(k-\frac q2\right).
    \label{eq:app_bubble_def}
\end{equation}
Thus
\begin{equation}
    S^{(2)}[W]
    =
    \frac{\gamma^2}{2}
    \int_q
    \left[
        C(q)^{-1}-\Pi(q)
    \right]
    \Str(W_qW_{-q}).
    \label{eq:app_W_kernel_raw}
\end{equation}
The masslessness of the $q=0$ angular mode follows from subtracting the equations for the two components of Eq. \eqref{eq:app_saddle}, which leads to
\begin{equation}
    R_\star^R-R_\star^A
    =
    C(0)\int_k
    \left[
        G_\star^R(k)-G_\star^A(k)
    \right].
\end{equation}
Using $G_\star^R(k)-G_\star^A(k) = (R_\star^R-R_\star^A) G_\star^R(k)G_\star^A(k)$, one obtains the identity
\begin{equation}
    C(0)\Pi(0)=1.
    \label{eq:app_Ward_identity}
\end{equation}
Hence \eqref{eq:app_W_kernel_raw} can be
written in the manifestly massless form
\begin{equation}
    S^{(2)}[W]
    =
    \frac{\gamma^2}{2}
    \int_q
    \mathcal K(q)\,
    \Str(W_qW_{-q}),
\end{equation}
where $\mathcal{K}(q)$ is given by Eq. \eqref{eq:LR3_kernel_split} and satisfies $\mathcal K(0)=0$, as required by invariance
under uniform rotations of the saddle. For $q\neq0$, $Q_q\simeq\Lambda W_q$ leads to 
\begin{equation}
    \Str(Q_qQ_{-q})
    =
    -\Str(W_qW_{-q})+O(W^3),
\end{equation}
thus retrieving the $\omega$-independent part of Eq. \eqref{eq:LR3_NLSM_general}.

We now extract the long-wavelength behavior of the two terms in
\eqref{eq:LR3_kernel_split}. Since
\begin{equation}
    C(r)\sim \frac{a_C}{|r|^{1+\sigma}},
\end{equation}
one has, for $0<\sigma<2$,
\begin{equation}
    C(q)=C(0)-a_\sigma |q|^\sigma+\cdots ,
\label{eq:Cq}
\end{equation}
where $a_\sigma = 2a_C\int_0^\infty dx\,\frac{1-\cos x}{x^{1+\sigma}}>0$.
Therefore
\begin{equation}
    C(q)^{-1}-C(0)^{-1}
    =
    \frac{a_\sigma}{C(0)^2}|q|^\sigma+\cdots .
    \label{eq:app_Cinv_scaling}
\end{equation}
On the other hand, the bubble contribution has a different -- subleading -- scaling in the small-$q$ limit. Since $\bar t(r)\sim r^{-(1+\sigma)}$, the deterministic dispersion has a small-momentum expansion analogous to \eqref{eq:Cq},
\begin{equation}
    \varepsilon(k)
    =
    \varepsilon(0)-A_\sigma |k|^{\sigma}+\cdots .
\end{equation}
Away from band-edge singularities,
\begin{equation}
    G_\star^{R/A}(k)
    =
    g_0^{R/A}
    +
    g_1^{R/A}|k|^{\sigma}
    +\cdots .
\end{equation}
In the difference $\Pi(0)-\Pi(q)$, the terms linear in the nonanalytic part
drop out by momentum-shift invariance. The leading nonanalytic contribution therefore comes
from the product of the two nonanalytic pieces,
\begin{equation}
    \Pi(0)-\Pi(q)
    \sim
    \int dk\,
    \left[
        |k|^{2\sigma}
        -
        \left|k+\frac q2\right|^{\sigma}
        \left|k-\frac q2\right|^{\sigma}
    \right].
    \label{eq:app_bubble_scaling_integral}
\end{equation}
The scaling of this integral is obtained by comparing the regions $k\sim q$
and $k\gg q$. In the first region, setting $k=qx$ gives immediately
\begin{equation}
    I_{k\sim q}(q)
    \sim
    |q|^{1+2\sigma}.
\end{equation}
In the second region, $k\gg q$, one may expand the integrand as
\begin{equation}
    |k|^{2\sigma}
    -
    \left|k+\frac q2\right|^{\sigma}
    \left|k-\frac q2\right|^{\sigma}
    \sim
    q^2 |k|^{2\sigma-2}.
\end{equation}
Thus
\begin{equation}
    I_{k\gg q}(q)
    \sim
    q^2 \int_{|q|}^{\Lambda} dk\, |k|^{2\sigma-2}.
\end{equation}
For $0<\sigma<1/2$, this integral is dominated by its lower limit
$k\sim q$, giving again $I_{k\gg q}(q)\sim|q|^{1+2\sigma}$. For $\sigma>1/2$, instead, the integral is dominated by the upper cutoff and
produces an analytic contribution $\sim q^2$. Since in this regime
$1+2\sigma>2$, the analytic $q^2$ term is the leading one. Therefore
\begin{equation}
    \Pi(0)-\Pi(q)
    \sim
    \begin{cases}
        |q|^{1+2\sigma}, & 0<\sigma<1/2,\\[1mm]
        q^2, & \sigma>1/2,
    \end{cases}
\end{equation}
up to logarithmic corrections at $\sigma=1/2$.

Finally, we restore the frequency. Let
\begin{equation}
    \Omega=\frac{\omega+i0}{2}.
\end{equation}
Expanding the logarithm in \eqref{eq:HS_rewritten} to first order in $\Omega$ gives
\begin{equation}
    \delta_\omega S
    =
    \Omega
    \sum_i
    \Str\left[
        G_{ii}[Q]\Lambda
    \right].
\end{equation}
To leading order in the long-wavelength expansion, this local term can be
evaluated for a locally uniform configuration. The saddle equation \eqref{eq:app_saddle_point} then gives
\begin{equation}
    R_i=C(0)G_{ii}[Q]
    \; \implies \;
    \delta_\omega S
    =
    \frac{\Omega}{C(0)}
    \sum_i
    \Str(R_i\Lambda).
\end{equation}
Using $\Str\Lambda=0$, this becomes
\begin{equation}
    \delta_\omega S
    =
    -i
    \frac{\Omega\gamma}{C(0)}
    \sum_i
    \Str(Q_i\Lambda).
\end{equation}
The saddle-level density of states is
\begin{equation}
    \nu(E)
    =
    -\frac{1}{\pi}\Im\int_k G_\star^R(k)
    =
    \frac{\gamma}{\pi C(0)}.
\end{equation}
Therefore
\begin{equation}
    S_\omega[Q]
    =
    -i
    \frac{\pi\nu(E)}{2}
    (\omega+i0)
    \sum_i
    \Str(\Lambda Q_i).
\end{equation}

\end{document}